\documentstyle[fleqn,10pt]{article}
\newcommand{\be}{\begin{equation}}
\newcommand{\ee}{\end{equation}}
\newcommand{\bi}[1]{\vspace{-3mm}  \bibitem{#1}}

\begin{document}

\begin{center}
{\large \bf STRINGS AND DISSIPATIVE MECHANICS. } \\
\vskip 4 mm

Vasily E. Tarasov \\
Nuclear Physics Institute, Moscow State University,
119899 Moscow, Russia \\
E-mail: tarasov@theory.npi.msu.su

\end{center}

Abstract:

Noncritical strings in the "coupling constant" phase space and bosonic
string in the affine-metric curved space are dissipative systems.
But the quantum descriptions of the dissipative systems have well known
ambiguities. We suggest some approach to solve the problems of
this description.  The generalized Poisson algebra for dissipative
systems is considered.

\section{ Introduction.}

The dissipative models in string theory are expected to have more broad
range of application:

1) Noncritical strings are dissipative systems in the "coupling constant"
phase space \cite{EMN,Nanop}. In this case, dissipative forces are defined
by non-vanishing beta-functions of corresponding coupling constant and by
Zamolodchikov metric \cite{EMN,Nanop}.

2) Problems of quantum description of black holes on the two-dimensional
string surface lead to the necessity of generalization of von Neumann
equation for dissipative systems \cite{EHNS,EMN,Nanop}.

3) The motion of a string (particle) in affine-metric curved space
is equivalent to the motion of the string (particle) subjected to
dissipative forces on Riemannian manifold \cite{Tarpl,Tarmpl}.
So the consistent theory of the string in the affine-metric curved
space is a quantum dissipative theory \cite{Tarpl,Tartmf2,Taryf,Tarmpl}.

But the quantum descriptions of the dissipative
systems (particles, strings,...) have well known problems
\cite{Lem,Hav1,Hav2,Edv,Hojm,Mess,Ber2,Kor2}.
The simple example of these problems is following.
It is easy to see that quantum equation of motion for dissipative
systems are not compatible with Heisenberg algebra. Let us consider
the quantum Langevin equation \cite{Haken}
\[ \dot a = (- \imath \alpha - \beta ) a + f(t) \ , \quad
 \dot a^{\dagger} = (- \imath \alpha - \beta) a^{\dagger} + f(t)^{\dagger} \]
We have
$ [\dot a, a^{\dagger}] + [a, \dot a^{\dagger}] \ = - 2 \beta $
On the other hand, the total time derivative of the Heisenberg algebra
\[ [a, a] \ = \ [a^{\dagger}, a^{\dagger}] \ = \ 0 , \
[ a, a^{\dagger}] = 1 \]
and Leibnitz rule lead to the following
$ [\dot a, a^{\dagger} ] + [a, \dot a^{\dagger}] \ = 0  $
So quantum dissipative equation of motion are not compatible with
canonical commutation relations and Heisenberg algebra.

Let us describe briefly some approaches to the quantum description of the
dissipative systems.

{\bf 1.1. Problems of canonical quantization.}

A) Lemos \cite{Lem} proved that canonical quantization relations is not
compatible with equation of motion for dissipative systems.
Note that Lemos considered the total time derivatives of the commutation
relations for the coordinates and momentums, used the Jacobi
identity and the dissipative equations of motion of Heisenberg
operator.

B) As is known that the equation of motion is the Euler-Lagrange equation
based on local a Lagrangian function when the Helmholtz conditions are
satisfied. Havas \cite{Hav2} considered a
general theory of multipliers which allows (by using the Helmholtz
conditions) a Lagrangian formulation for a broad class of the equation
of motion of the dissipative systems, which can not
fit into Lagrangian mechanics by usual approach.
Havas hence noted that the quantization of systems described by
Lagrangian of the above type is either impossible or ambiguous \cite{Hav1}.
This is follows from the fact that in classical mechanics of the
dissipative systems there are many quite different Lagrangians and
Hamiltonians leading to the same equations of motion  \cite{PhysRep}.
So we  do not know which of the possible Lagrangians is corrected
and one to choose for quantization procedure.

C) Edvards \cite{Edv} showed that, although classical Hamiltonians are
necessary for canonical quantization, their existence is not sufficient
for it. The quantization of the Hamiltonian which is not canonically
related to the energy is ambiguous and therefore the results are
conflicting with physical interpretations.
It is not sufficient for the
Hamiltonian to generate the equation of motion, but Hamiltonian must
also be necessarily related via canonical transformation to the total energy
of the system. However, this condition can only be met by
conservative systems, thus excluding dissipative systems from
possible canonical quantization \cite{Edv}.

D) Hojman and Shepley \cite{Hojm} started with classical equation of motion
and set very general quantization conditions (relation that the coordinate
operators commute). The total time derivative of this commutation
relation was considered and the commutator of the coordinate operator
and the velocity operator form a symmetric tensor operator was
showed. They proved that classical analog of this tensor operator is
a matrix which inverse matrix satisfies the Helmholtz conditions.
Using the Jacobi identity for the coordinate and the velocity, Hojman and
Shepley conclude the following:
the general quantization condition implies that the equation of motions
are equivalent to the Euler-Lagrange equations with some Lagrangian.

{\bf 1.2. Generalization of von Neumann equation.}

The generalized von Neumann equations proposed until now
\cite{Mess,Ber2,EHNS,Kor2}, which should describe dissipative
and irreversible processes are derived by the addition the
superoperator which acts on the statistical operator and describes
dissipative part of time evolution. Note that proposed generalizations
of the von Neumann equation are derived heuristically. For a given set
of generalized equations different requirements for superoperator
exist. Most of the requirements proposed until now, which should
determine superoperator uniquely are not unique themselves and so one
has to deal with the problems arising from these ambiguities.
The superoperator form is not determine uniquely.
Moreover, the generalizations of the von Neumann equation are not
connected with classical Liouville equation for dissipative systems
\cite{Steb,Prig2,Fron}. Hence the quantum description of
the dissipative system dynamics used the generalizations of von Neumann
equation proposed until now is ambiguous.

{\bf 1.3. Nonassociative Lie-admissible quantization.}

The generalization of the canonical quantization of the dissipative systems
was proposed by Santilli \cite{Sant1}. Santilli
showed that the time evolution law of dissipative equation
not only violates Lie algebra law but actually does not characterize an
algebra. Therefore Santilli suggested,  as a necessary condition
to preserve the algebraic structure, that the  quantum dynamics of the
dissipative systems should be constructed within the framework of the
nonassociative algebras. This is exactly the case of the noncanonical
quantization at the level of the nonassociative Lie-admissible (Lie-isotopic)
enveloping algebra worked out by Santtili.

The quantization of the dissipative systems was proposed by Santilli
\cite{Sant1} as an operator image of the
Hamiltonian-admissible and Birkhoffian generalization of the
classical Hamiltonian mechanics. The generalized variations used by
Santilli \cite{Sant4} to consider the dissipative processes in the field of
the holonomic variational principles are connected with the generalized
multipliers suggested by Havas \cite{Hav2} and therefore lead to
an ambiguity in generalized variations.

{\bf 1.4. Nonholonomic variational principle. }

Sedov \cite{Sed1}- \cite{Sdch1} suggested
the variational principle which is the generalization of the least
action principle for the dissipative and irreversible processes. The
holonomic and nonholonomic functionals are used to include the dissipative
processes in the field of the variational principle.

Nonholonomic principle was suggested in \cite{Taryf,Tartmf1,Tarictp} to
generalize the classical mechanics in phase space.
The suggested form of classical mechanics of the dissipative systems
in the phase space can be used to consider the generalizations of
canonical quantization for dissipative systems and von Neumann equation.

{\bf 1.5. We can conclude the following:}

 1. Canonical quantization of the dissipative systems are impossible if
all of operators in quantum theory are associative \cite{Hojm,Lem}.

2. Equation of motion  of dissipative systems are not
compatible with canonical commutation relations (with Heisenberg algebra).

3. Coordinate and momentum operators must be satisfy the canonical
commutation relations. The generalized operator algebra must not violate
Heisenberg algebra.

4. Generalization of the von Neumann equation must be connected with
classical Liouville equation for dissipative systems \cite{Tartmf1,Tarictp}.

5. Hamiltonian must be canonically related to the physical energy $(T+U)$
of the dissipative system \cite{Edv}.

6. Total time derivative of the dissipative system operator does not satisfy
the Leibnitz rule \cite{Brat1}.

7. Canonical quantization of the dissipative systems described within
framework of holonomic variational principles (least action principle)
is either impossible or ambiguous \cite{Hav1,Hav2}.

8. Quantum description of dissipative systems within framework of quantum
kinetics is very popular and successful, but it is not valid in the
fundamental theories such as string theory.

9. Dissipative systems can be described within framework of
the nonholonomic variational principle \cite{Sed1} - \cite{Sdch1}.

In this paper we consider the some main points of the quantum description
of the dissipative systems which take into account these conclusions.

{\bf 1.6. }

In order to solve the problems of the quantum description of dissipative
systems we suggest to introduce an operator W in addition to usual
(associative) operators.
{\it The suggested operator algebra does not violate Heisenberg algebra
because we extend the canonical commutation relations by introducing
an operators of the nonholonomic quantity in addition to the usual
(associative) operators of usual (holonomic) coordinate-momentum functions.
That is the coordinate and momentum satisfy the canonical commutation
relations.} To satisfy the generalized commutation relations the operator
$W$ of nonholonomic quantity must be nonassociative non-Lieble (does not
satisfied the Jacobi identity) operator
\cite{Taryf,Tartmf2}. As the result of these properties the total
time derivative of the multiplication and commutator of the operators
does not satisfies the Leibnitz rule.
This lead to compatibility of quantum equations of motion for dissipative
systems and canonical commutation relations.
The suggested generalization of the von Neumann equation is connected
with classical Liouville equation for dissipative systems.

\section{ Classical Dissipative Mechanics.}

{\bf 2.1. Generalization of the Poisson brackets.}

Let the coordinates $ z^k , \ ( k =1,..., 2n), \ $ where
$ \ z^i=q^i, \ z^{n+i}=p_i \ \ (i=1,...,n) \ $ and $ \ w, t $ of the
(2n+2)-dimensional extended phase space be connected by the equations
\be
\label{4}
d w \ - \ a_k (z,t) \ d z^k \  = \ 0
\ee
where $ a_k \ ( k=1,...,2n) $ are the vector functions in phase space.
Let us call the dependence $w$ on the coordinate $q$ and momentum $p$ the
holonomic-nonholonomic function and denote $ w=w(z) \in F^* (M) $
(generalization of the potential for closed and nonclosed 1-forms).
If the vector functions satisfy the relation
\be
\label{5}
\frac{\partial a_k (z) }{\partial z^l} = \frac{\partial a^l
(z)}{\partial z^k }
\ee
where k;l= 1,...,2n , the coordinate $w$ is the holonomic function ($ w
\in F (M) = \Lambda^0 (M) $), i.e. potential for closed 1-form.
By definition, if these vector functions don't satisfy the
relation (\ref{5}) the object $w(z)$ we call the nonholonomic function
($ w \in F^* / F $) -- generalized potential of the
nonclosed 1-form $d (d w) \not \equiv 0$.

Let us define the generalized Poisson brackets for the generalized
potentials $f,g,s \in F^* $ of the (closed and nonclosed) 1-forms
$ \alpha= df, \beta=dg, \gamma=ds \ (d^2 f \not \equiv 0  \ f \in F^*)$
on the symplectic manifold $(M,\omega)$
\be
[f,g] \ \equiv \Psi(\alpha, \beta) \ = \ \omega (X_{\alpha}, X_{\beta})
\ = \ \Psi^{kl} a_k b_l
\ee
where $ X_{\alpha}: \ i(X_{\alpha}) \omega =  \alpha \ ; \ $
 $ \omega  $ - closed ($d \omega = 0$) 2-form \cite{Godb,Maslov},
called symplectic form; $i$ --  internal multiplication of
the vector fields and the form \cite{Godb}.
$ \Psi^{kl}$  -- contrvariant 2-tensor, which is the matrix
inverse to matrix of the symplectic form
and satisfies \cite{Kirilov,Maslov}:

a)  Skew-symmetry: $ \qquad \ \   \Psi^{kl} =  \Psi^{lk} $

b)  Zero Schouten brackets:
$  [ \Psi, \Psi]^{slk} =  \Psi^{sm} \partial_m \Psi^{lk} +
 \Psi^{lm} \partial_m \Psi^{ks} +  \Psi^{km}  \partial_m \Psi^{sl} = 0  $

The generalized Poisson brackets can be represented by
\be
[f,g] \equiv \frac{\delta f }{\delta q^i}\frac{\delta
g }{\delta p_i} - \frac{\delta f }{\delta p_i} \frac{\delta g}{\delta q^i}
= a_i b_{n+i} - a_{n+i} b_i
\ee
where
\[ \alpha = a_k (z) dz^k = a_i dq^i  + a_{n+i} dp^i \ ,
\qquad  \beta = b_k (z) dz^k = b_i dq^i  + b_{n+i} dp^i \]

The basic properties of the generalized Poisson brackets:

$ 1) Skew-symmetry:  \quad \forall f,g \in F^* \qquad
[f,g] = - [g,f] \in F ; $

$ 2) Jacobi \ identity:  \quad \forall f, g, s \in F \qquad
J[f,g,s] = 0; $

$ 3) Nonliebility:  \quad \forall f,g,s, \in F^* : \ f \vee g \vee s
\in F^* / F \qquad  J[f,g,s] \not \equiv 0; $

$ 4) Distributive \ rule:  \forall f,g,s \in F^* \quad
[ \alpha f + \beta g ,s ] = \alpha [f,s] + \beta [ g ,s ] $

$ 5) Leibnitz \ rule:  \quad \forall f,g \in F^* \quad
\frac{ \partial }{ \partial t } [f, g] = [ \frac{ \partial }{ \partial t }
  f, g] + [f ,  \frac{ \partial }{ \partial t }  g]; $

where $ \ J[f,g,s] \equiv [f,[g,s]] + [g,[s,f]] + [s,[f,g]]  , \  $

$\alpha \ $  and $ \ \beta $ are the real numbers.

If this bilinear operation "generalized Poisson bracket"
is defined on the space of generalized potentials $ F^* (M)$ ,
then the manifold $M$ is called poisson
manifold, and the space  $ F^* (M)$ -- generalized Poisson algebra $P^*_0$.
Generalized Poisson algebra $P^*_0$ is not Lie algebra:
Jacobi identity for $F^* / F$ is not satisfied.
But Jacobi identity is satisfied on the space $F(M)$. For this reason
we can define in the space $F(M)$ a Lie algebra, which is the
Poisson algebra $P_0$ \cite{Maslov}.
It is easy to verify that this properties of the generalized Poisson
brackets for the holonomic functions coincide with the properties of
the usual Poisson brackets \cite{Lanc}.

That is generalized Poisson algebra $P^*_0$ contain a subalgebra
which is the usual Poisson algebra $P_0$. Generalized Poisson
bracket is the holonomic function so this subalgebra $P_0$
is the ideal of the algebra $P_0^*$

{\bf 2.2.} Let us consider now the characteristic properties of the
physical quantities:

$ 1) \quad [p_i ,p_j ] = [q^i ,q^j ] = 0 \qquad and
\qquad [q^i ,p_j ] = \delta^i_j $

$ 2) \quad [w, p_i ] = w^q_i \qquad and
  \qquad [w, q^i ] = - w_p^i  \qquad i \not \equiv j , \qquad [w, w] = 0 $

$ 3) \quad [[w,p_i ],p_j ] \not \equiv [[w,p_j ], p_i ]
\quad or \quad  J[p_i ,w,p_j ] = \omega_{ij} \not \equiv 0
 \qquad i \not \equiv j $

$ 4) \quad [[w, q^i ], q^j ] \not \equiv [[w, q^j ], q^i ]
\quad or \quad  J[q^i ,w,q^j ] = \omega^{ij} \not \equiv 0
 \qquad i \not \equiv j $

$ 5) \quad [[w, q^i ], p_j ] \not \equiv [[w, p_j ], q^i ]
\quad or \quad  J[q^i ,w,p_j ] = \omega^i_j \not\equiv 0 $

where
\be
\omega^i_j \ \equiv \ \frac{\partial w^q_j}{\partial p_i} -
\frac{ \partial w_p^i }{ \partial q^j } \ =  \
 \frac{\delta^2 w}{\delta p_i \delta q^j} - \frac{\delta^2 w
}{\delta q^j \partial p_i}
\ee
This object $ \omega^{kl} \ (k,l=1,...,2n) \ $ characterizes
deviation from the condition of integrability (\ref{5}) for
the equation (\ref{4}) and by the Stokes theorem
\be
\oint_{\partial M} \delta w =  \int_{M} \omega^{kl} \ dz^k \wedge dz^l
 \not \equiv 0
\ee
Note that $w$ is the nonholonomic object if one of  $\omega^{kl}$ is not
trivial.
Therefore some of the properties 3-5 can be not satisfied but one of
it must be carry out if we consider the dissipative processes.

{\bf 2.3. Equation of motion for dissipative systems.}

Equations of motion for dissipative systems in the phase space is given by
\be
\label{3}
\frac{dq^i}{dt} =  \frac{\delta h}{\delta p_i} -  w_p^i
= \frac{\delta(h-w)}{\delta p_i} \quad ; \qquad
\frac{dp_i}{dt} = - \frac{\delta h }{\delta q^i} + w^q_i =
 - \frac{\delta(h-w) }{\delta q^i}
\ee
where  $ \delta w(q,p)=  w^q_i \delta q^i + w_p^i \delta p_i $

If we take into account generalized Poisson brackets the equation of
motion in phase space for dissipative systems (\ref{3}) takes the form
\be
\label{6}
\frac{dq^i}{dt} = [q^i,h-w] \qquad \frac{dp_i}{dt} = [p_i, h-w]
\ee
The total time derivative of the physical quantity $A = A(q,p,t) \in F$
is given by
\be
\label{7}
\frac{dA(q,p,t)}{dt} = \frac{\partial A(q,p,t)}{\partial t } + [A, h - w]
\ee
The equation of motion (\ref{6}) can be derived from the
equation (\ref{7}) as a particular case. Note that any term which added to
the Hamiltonian $h$ and nonholonomic object $w$ does not change
the equations of
motions (\ref{6}), (\ref{7}). This ambiguity in the definition of the
Hamiltonian is easy to avoid by the requirement that Hamiltonian must
be canonically related to the physical energy of the system \cite{Edv}:
$ \ [w,q^i] \ = \ 0 $

It is easy to see that total time derivative of the generalized
Poisson brackets does not satisfies the Leibnitz rule
\be
\frac{d}{dt} [f, g] = [ \frac{d}{d t } f, g] + [f , \frac{d}{dt} g]
 + J[f,w,g]
\ee

{\bf 2.4. Generalized Poisson algebra of 1-forms.}

It is known that Poisson brackets can be defined for nonclosed
differential 1-forms $\alpha = a_k (z) d z^k $ on the symplectic manifold
$(M,\omega)$ \cite{Maslov}, where $\omega$ - closed ($d \omega = 0$)
2-form \cite{Arnold}. Poisson bracket for two 1-forms
$ \alpha = a_k (z) d z^k  $ and $\beta =  b_k (z) d z^k $ is 1-form
 $(\alpha, \beta)$, defined by
\be
(\alpha, \beta) \ = \ d \Psi (\alpha, \beta) \ + \ \Psi (d \alpha, \beta) \
+ \ \Psi(\alpha, d \beta)
\ee
that is the map of 1-forms
$ \Lambda^1 (M) \times \Lambda^1 (M) \rightarrow  \Lambda^1 (M)$

If this bilinear operation "Poisson bracket" is defined on the space
of 1-forms  $ \Lambda^1 (M)$ , then the manifold $M$ is called poisson
manifold, and the space  $ \Lambda^1 (M)$ -- Poisson algebra $P_1$.
Poisson algebra $P_1$ is a Lie algebra. It is caused by skew-symmetry
$(\alpha, \beta) = - (\beta, \alpha)$ and Jacobi identity:

$ ((\alpha, \beta), \gamma) + ((\beta , \gamma) \alpha) +
((\gamma , \alpha), \beta) \ = \ 0 $

In order to describe dissipative systems we suggest to generalize the
Poisson algebra $P_1$. Let us define the bilinear operation on
$ \Lambda^1 (M)$ :
Generalized Poisson bracket of two  1-forms $\alpha$ and $\beta$ is
1-form $(\alpha, \beta)$, defined by
\be
\label{pa}
[ \alpha, \beta ] \ = \ d ( \ \Psi (\alpha, \beta) \ )
\ee
It is easy to see the Jacobi identity for non-closed 1-forms
is not satisfied. Therefore generalized Poisson algebra $P^{*}_1$
is not Lie algebra.
But Jacobi identity is satisfied for closed 1-forms. For this reason
closed 1-forms define a Lie algebra, which is the Poisson algebra  $P_1$.
So generalized Poisson algebra $P^*_1$ contain a subalgebra
which is the usual Poisson algebra $P_1$. Generalized Poisson
bracket is the closed 1-form so this subalgebra $P_1$
is the ideal of the algebra $P_1^*$ and the exact algebraic
diagram exists:

$ 0 \rightarrow  P_1  \rightarrow P_1^*  \rightarrow  P_1^* / P_1
  \rightarrow  0 $

Generalized Poisson bracket for 1-forms satisfy the properties:

1)  Skew-symmetry: $ \quad \forall \alpha, \beta \in  P_1^*  \qquad
[\alpha, \beta] = - [\beta, \alpha]
\in  P_1 ;$

2) Jacobi \ identity: $ \quad \forall \alpha, \beta, \gamma \in  P_1 \qquad
J[\alpha, \beta , \gamma] = 0; $

3)  Nonliebility:: $ \quad \forall \alpha, \beta, \gamma \in  P_1^*  :
\ \alpha \vee \beta \vee \gamma
\in  P_1^* /  P_1 \qquad  J[\alpha , \beta , \gamma] \not \equiv 0; $

4)  Distributive \ rule: $ \forall \alpha, \beta, \gamma \in  P_1^*  \quad
[ a \alpha + b \beta , \gamma ] = a [\alpha , \gamma] + b [ \beta , \gamma]
\ , $

where
$ J[\alpha , \beta , \gamma] \equiv
[\alpha, [\beta , \gamma]] + [ \beta ,[\gamma, \alpha]] +
[\gamma, [\beta, \alpha]] \ , \ \ \ $
$ a \ $ and $ \ b $  are the real numbers.

That is the structure of the anticommutative nonassociative
algebra, which is not Lie algebra, is naturally defined
in the space of all 1-forms $ \Lambda^1 (M)$ on the symplectic manifold
$M$. This algebra is a generalization of anticommutative
nonassociative Lie algebra of closed 1-forms and  contain a subalgebra
(ideal) which is this Lie algebra.

{\bf 2.5. Liouville equation for dissipative systems.}

It is easy to obtain the dissipative analogue of the
Liouville equation \cite{Steb,Fron,Prig2,Tartmf2,Tarictp}:
\be
\label{10}
\frac{d}{dt}  \rho (q,p,t) \ = \ - \ \omega (t,q,p) \rho (q,p,t)
\ \ or \quad
\imath \frac{\partial}{\partial t}  \rho (q,p,t) \ = \ \Phi L \rho (q,p,t)
\ee
where $ \omega \ = \ \sum_{i=1}^{n} \omega^i_i \
= \ \sum_{i=1}^{n} \ J[q^i, W, p_i]  \ \ and \quad
\Phi = \imath ( \ \frac{\delta (h-w) }{\delta q^k}
\frac{\partial}{\partial p_k} - \frac{\delta (h-w)}{\delta p_k}
\frac{\partial}{\partial q^k} - \omega (t,q,p) \ ) $
is generalization of the Liouville operator \cite{Prig1}.
In addition to the Poincare-Misra theorem \cite{Prig1} can
be obtained the statement:
"There exists the Liapunov function of the coordinate and momentum
in the dissipative Hamiltonian mechanics". Let us define the
function $ \eta (q,p,t) \equiv -  ln \rho (q,p,t) $ and assume
$ \omega > 0 $. The equation (\ref{10}) shows that
\[ d \eta (q,p,t) / dt \ = \ \omega (t,q,p) \]
and the function $\eta$ satisfies the relations $ {d \eta}/{dt} > 0 $.

\section{Quantum Dissipative Mechanics.}

{\bf 3.1. Generalization of the canonical commutation relations.}

In order to solve the problems of the quantum description of dissipative
systems we suggest to introduce an operator W in addition to usual
(associative) operators.
Let us use the usual rule of definition of the quantum physical
quantities which have the classical analogues: If we consider
the operators A,B,C of the physical quantities a,b,c  which
satisfy the classical Poisson brackets $ [a,b] = c $, then
the operators must satisfy the relation: $ [A,B] \equiv (AB) - (BA) =
\imath \hbar C $. If we take into account the characteristic
properties the physical quantities operators are defined by the
following relations:

$ 1) \quad [ Q^i, Q^j ] = [ P_i, P_j ] = 0 \qquad [ Q^i, P_j ] =
\imath \hbar \delta^i_j $

$ 2) \quad [ W, P_i ] = \imath \hbar W^q_i \qquad [ W, Q^i ] = -
\imath \hbar W_p^i  \qquad [W, W] = 0 $

$ 3) \quad [[W, P_i], P_j] \not \equiv [[W, P_j], P_i] \quad i \not \equiv j
\quad or \quad J[P_i, W, P_j ] = \Omega_{ij} \not \equiv 0 $

$ 4) \quad [[W, Q^i], Q^j] \not\equiv [[W, Q^j], Q^i] \quad i \not \equiv j
\quad or \quad J[Q^i, W, Q^j ] = \Omega^{ij}  \not \equiv 0 $

$ 5) \quad [Q^i ,[W, P_j ]] \not \equiv [P_j, [W, Q^i]] \quad or \quad
J[Q^i, W, P_j ] = \Omega^i_j \not \equiv 0 $

where
$ J[A,B,C] = -{1}/(\hbar^2) \ ( \ [ A [B C]] + [B [C A]] + [C
[A B]] \ ) $

and $ \ Q^{\dagger} = Q ; \ P^{\dagger} = P; \  W^{\dagger} = W \ $.
{\it Let us require that the canonical quantum commutation rules be
a part of this rules.} To satisfy the commutation relations and canonical
commutation rules for the operator of the holonomic function the operators
of the nonholonomic quantities must be nonassociative. It is sufficient
to require that the operator W satisfies the following conditions:

 1) \quad  left \ and \ right \ associativity:
$ (Z^k,Z^l,W) =(W,Z^k,Z^l) = 0 $

 2) \quad left-right \ nonassociativity:
$ (Z^k,W,Z^l) \not \equiv 0  \quad if \quad k \not \equiv l $

where $ k;l = 1,..,2n ; \ Z^i=Q^i \ $  and $ \ Z^{n+i}=P_i ; \ i=1,...,n ; $
$ (A,B,C) \ \equiv \ (A(BC)) - ((AB)C) \ $ called associator.

{\bf 3.2. Quantum equation of motion for dissipative systems. }

The state in the quantum dissipative mechanics can be represented
by the "matrix-density" (statistical) operator $ \rho (t) $ which
satisfy the condition $ \rho^{\dagger} (t) = \rho (t) $.
The time variations of the operator of physical quantity $ A(t)
\equiv A(Q,P,t) $ and of the operator of state $ \rho (t) $
are written in the form
\be
\label{11}
\frac{dA}{dt} = \frac{\partial A}{\partial t} +
 \frac{\imath}{\hbar} [ H-W, A ]
\ee
\be
\label{12}
\frac{d \rho }{dt} = - \frac{1}{2} [ \ \rho \ , \ \Omega \ ]_{+}
\qquad
\frac{\partial }{\partial t} \rho \  = \
\frac{\imath}{\hbar} [ \rho, H ] \ + \
\frac{\imath}{\hbar} [ W, \rho ] \ -  \
\frac{1}{2} [ \ \rho \ , \ \Omega \ ]_{+}
\ee
where anticommutator $ [ \  , \  ]_{+} $ is the consequence of the
hermiticity for the density operator $\rho $ and for the operator
$ \Omega $, which is defined by
\[ \Omega \ = \ \sum_{i=1}^{n} \Omega^i_i \
= \ \sum_{i=1}^{n} \ J[Q^i, W, P_i]  \]

The solution of the first equation may be written in the form
\be
 A(t) = S(t,t_0) A(t_0) S^{\dagger} (t,t_0)
\quad where \qquad
S(t,t_0) = Texp \frac{\imath}{\hbar} \int^{t}_{t_0}d \tau \ (H-W)(\tau)
\ee
T-exponent is defined as usual, but we must take into account the
following flow chart
\[ exp \ A = 1 + A + \frac{1}{2} (AA) + \frac{1}{6} ((AA)A) + \frac{1}{24}
(((AA)A)A) + ...  \ \ . \]

The solution of the equation (\ref{12}) is given by
\be
\rho (t) = U(t,t_0) \rho (t_0) U^{\dagger} (t,t_0) \quad where \quad
U(t,t_0) = Texp \frac{1}{2} \int^{t}_{t_o} d \tau \ \Omega ( \tau )
\ee
In this way the time evolution of the physical quantity operator is
unitary and the evolution of the state operator is nonunitary.
It is easy to verify that the pure state  at the moment
$ t = t_0 $ ( $ \rho^{2} (t_0) = \rho (t_0) $ )
is not a pure state at the next time moment $ t \not \equiv t_0 $. We can
define the entropy operator $ \eta $ of the state $ \rho (t) $   :
$ \ \eta (t) = - \ ln \rho (t) $. The entropy operator satisfies the
equation $ d \eta (t) / dt \  = \ \Omega $

{\bf 3.3. Generalized Leibnitz rule.}

It is easy to see that the commutator with nonassociative operator $W$
and the total time derivative of both the quantum Poisson brackets
and of the multiplication of the two operators do not satisfy the
Leibnitz rule
\be
[AB, W] =  A [  B, W] + [A , W ] B \ + \ (A,W,B)
\ee
\be
\frac{d}{dt} [A, B] = [ \frac{d}{d t } A, B] + [A , \frac{d}{dt} B]
\ + \ J[A,W,B]
\ee
\be
\frac{d}{dt} (A B) = (( \frac{d}{d t } A) B) + (A ( \frac{d}{dt} B))
\ + \ (A,W,B)
\ee
where A and B are the associative operators (operators of the
holonomic functions).
This lead to compatibility of quantum equations of motion for dissipative
systems and canonical commutation relations.
We can formulate the path integration and generating
functional in the quantum field theory  which was
considered in the papers \cite{Tar2,Tarpl,Tarmpl}.

{\bf 3.4. Some applications.}

The dissipative quantum scheme suggested in  \cite{Taryf,Tartmf2} and
considered in this paper allows to formulate the approach to the quantum
dissipative field theory. As an example of the dissipative quantum field
theory the sigma-model approach to the quantum string theory was
considered in the recent papers \cite{Tarpl,Tarmpl}.
Conformal anomaly of the energy momentum tensor trace for closed
bosonic string on the affine-metric manifold and two-loop metric
beta-function for two-dimensional nonlinear dissipative sigma-model
were calculated \cite{Taryf,Tarpl,Tarmpl}.

\end{document}